\documentclass[12pt]{article}
\usepackage[margin=2cm]{geometry}
\usepackage{comment}
\usepackage{amsmath,amssymb,extarrows,mathtools,graphicx,subfigure,setspace}
\usepackage{cite}
\usepackage{slashed}
\usepackage{color}
\makeatother

\newcommand{\be}{\begin{equation}}
\newcommand{\bea}{\begin{eqnarray}}
\newcommand{\eea}{\end{eqnarray}}
\newcommand{\ba}{\begin{array}}
\newcommand{\ea}{\end{array}}
\newcommand{\ee}{\end{equation}}
\newcommand{\bes}{\begin{equation*}}
\newcommand{\beas}{\begin{eqnarray*}}
\newcommand{\eeas}{\end{eqnarray*}}
\newcommand{\bas}{\begin{array*}}
\newcommand{\eas}{\end{array*}}
\newcommand{\ees}{\end{equation*}}
\newcommand{\nn}{\nonumber}
\numberwithin{equation}{section}
\begin{document}
\onehalfspacing
\noindent
\begin{titlepage}
\hfill
\vbox{
    \halign{#\hfil         \cr
           IPM/P-2012/030 \cr
                      } % end of \halign
      }  % end of \vbox
\vspace*{20mm}
\begin{center}
{\Large {\bf Holographic Aspects of Two-charged  Dilatonic Black Hole in AdS$_5$}\\
}

\vspace*{15mm}
\vspace*{1mm}
{Mohsen Alishahiha, M. Reza Mohammadi Mozaffar and  Ali Mollabashi }

 \vspace*{1cm}

{\it  School of physics, Institute for Research in Fundamental Sciences (IPM)\\
P.O. Box 19395-5531, Tehran, Iran \\ }

\vspace*{.4cm}

{E-mails: {\tt alishah, m$_{-}$mohammadi, mollabashi@ipm.ir}}%

\vspace*{2cm}
\end{center}
\begin{abstract}
We study certain features of a strongly coupled theory whose gravitational dual is given by two-charge dilatonic black hole in AdS$_5$ which has recently been used to study holographic Fermi liquids. By making use of the gravity description, we have studied conductivity, holographic entanglement entropy and dynamics of a charged scalar field. In particular at low energy we find that the temperature dependence of the real part of the conductivity goes as $T^3$ and the background is stable against scalar condensations.
\end{abstract}
\end{titlepage}
%%%%%%%%%%%%%%%%%%%
\section{Introduction}

AdS/CFT correspondence \cite{M:1997,GKP:1998,W:1998} has provided a framework to study strongly coupled field theory by making use of a classical gravity. In particular, there have been several attempts to apply the AdS/CFT correspondence in condensed matter physics (for reviews see \cite{SM:2008,H:2009,Herzog:2009,M:2009}). Since in this application we are  typically dealing with matters at finite temperature and density, the gravitational duals should be charged black holes (for early works see \cite{HKSS:2007}).  In this context, the gravity description has been used to explore, {\it e.g.} the existence of Fermi surfaces in the system and  the properties of the low energy excitations near the Fermi surface.

To construct the gravitational dual one may use an Einstein-Maxwell theory on a Reissner-Nordstr\"on black hole in asymptotically AdS geometry which, indeed, has been used to study certain features of strongly coupled systems at finite temperature and density \cite{Lee:2008xf}. More precisely, fermionic retarded Green's function in the dual theory can be calculated by solving the Dirac equation in the bulk. It is then possible to extract certain information such as the existence of Fermi surfaces and also the properties of the low energies excitation near the Fermi surface \cite{{Liu:2009dm},{Cubrovic:2009ye},{FLMV:2009},{FILMV:2011}}.

In rather more involved models, dilatonic black hole solutions in asymptotically AdS geometry have also been used to construct gravitational duals for fermionic systems at finite density such as non-Fermi liquids (see for example \cite{{GR:2009},{Goldstein:2009cv}}). In particular a two-charge dilatonic black hole in AdS$_5$ has been considered in
\cite{GR:2009} where it was shown that it might provide a suitable gravitational description for the Landau Fermi liquids. In particular, the model exhibits vanishing entropy and  its specific heat is linear in temperature at low energies. Moreover, using a Dirac fermion in the bulk it was shown that the background supports normal modes for the massless bulk fermions which is expected for a Fermi liquid \cite{{GR:2009},{GR:2012}}.

From five dimensional maximally gauged supergravity, this two-charge black hole can be obtained by setting two charges non-zero and equal while the third one is zero. Although the resultant geometry has a naked singularity in the extremal limit, the physical quantities may be computed by imposing suitable boundary conditions\footnote{Naked singularities are more disscused in \cite{VE:2002}.}.

The corresponding two-charge black-hole in an asymptotically AdS$_5$ geometry may be obtained from the following Lagrangian \cite{GR:2009} 
\bea \label{action}
\mathcal{L}=\frac{1}{2\kappa^2}\left[\mathcal{R}-\frac{1}{4}e^{4\alpha}(F_{MN})^2-12(\partial_M \alpha)^2+\frac{1}{R^2}(8e^{2\alpha}+4e^{-4\alpha})\right]
\eea
where $\alpha$ is  a neutral scalar which plays the role of dilaton. $M,N,\cdots$ indices are five dimensional space-time  indices. 

The two-charge black hole solution is\cite{GR:2009}\footnote{In what follows we set $R=1$.}
\bea\label{ds2}
&&ds^2=[\frac{r(r^2+Q^2)}{R^3}]^{2/3}\left[-\frac{(r^2-r_0^2)(r^2+r_0^2+2Q^2)}{(r^2+Q^2)^2}dt^2+d\vec{x}^2+\frac{R^4dr^2}{(r^2-r_0^2)(r^2+r_0^2+2Q^2)}\right],\cr
&&\cr
&&e^{6\alpha}=1+\frac{Q^2}{r^2},\;\;\;\;\;\;\;\;A_t(r)=-\frac{\sqrt{2}Q}{R}\left(1-\frac{r_0^2+Q^2}{r^2+Q^2}\right).
\eea

The Hawking temperature is $T=\frac{r_0}{\pi}$ and the extremal case corresponds to $r_0=0$ where the space-time develops a naked singularity. Although the geometry has 
naked singularity, there are several advantages to work with this dilatonic black hole\cite{GR:2012}. 

In fact, unlike the case of RN AdS back hole, in the case of two-charged dilatonic black hole, the exact position of the Fermi surfaces of massless fermions is known which makes it tractable to study their properties. More importantly the specific heat, entropy and the shear viscosity of the dilatonic black hole are proportional to the temperature. In particular the entropy of the IR geometry is zero. While it is believed that the back reaction of the fermionic matters could distort the RN AdS geometry \cite{Hartnoll:2009ns}, it is not the case for the dilatonic black hole case (for more details and explanations concerning these points see \cite{GR:2012})\footnote{See \cite{W:2011} for some numerical studies of the fermionic properties of this background.}.

These properties might be understood from the fact that unlike the RN AdS black hole, where the near horizon geometry in the extremal case develops an AdS$_2$ geometry, in the present case the near horizon geometry is conformally AdS$_2$. More precisely, taking the near horizon limit for the near extremal case one finds\footnote{To be precise  the near horizon limit is defined by taking $\lambda\rightarrow 0$ while keeping the following quantities fixed $$\xi=\frac{\lambda }{2r},\;\;\;\;\xi_0=\frac{\lambda }{2r_0},\;\;\;\;\;{\tau}=t{\lambda.}$$ We note also that in this limit the metric needs to be scaled by a proper power of $\lambda$.\label{foot}}
\bea\label{AdS2} 
ds^2=\left(\frac{1}{2Q\xi}\right)^{2/3}\left\{\frac{1}{2\xi^2}\left[-\left(1-\frac{\xi^2}{\xi_0^2}\right)d\tau^2
+\frac{d\xi^2}{\left(1-\frac{\xi^2}{\xi_0^2}\right)}\right]+Q^2d\vec{x}^2\right\},
\eea
which is conformally an AdS$_2$ geometry with vanishing entropy.

The aim of the present work is to further explore certain features of the theory dual to the dilatonic black hole. To be precise, by perturbing certain components of the gauge field in the bulk, we calculate the corresponding retarded Green's function of a conserved current which in turns can be used to read the conductivity. We will also compute the holographic entanglement entropy in this geometry. 

In order to further understand the low energy phase of the system we will also consider a charged scalar field in the two-charge dilatonic black hole which can be used to compute the corresponding retarded Green's function of a scalar operator in the dual theory. The resultant  retarded Green's function may be used to study  the possibility of instabilities due to scalar condensation and thus further  explore  the behavior of the system near the quantum critical points.

The paper is organized as follows. In the next section we will compute the DC conductivity of the model. In section three the holographic entanglement entropy is computed. In section four we shall study a scalar field in the geometry and the last section is devoted to conclusions. Some details are presented in the Appendix.

%%%%%%%%%%%%%%%%%%%%%%%%%%%%%%%%%%%%%%%%%%%%%%%%%%%%%%%%%%%%%%%%%%%%%%%%%%
%%%%%%%%%%%%%%%%%%%%%%%%%%%%%%%%%%%%%%%%%%%%%%%%%%%%%%%%%%%%%%%%%%%%%%%%%%

\section{Optical conductivity}
In this section we study the optical conductivity of a finite density system whose gravity dual is given by \eqref{ds2}. The local $U(1)$ gauge field in the bulk is dual to a conserved current of a global $U(1)$ symmetry in the dual theory. In terms of the conserved current, the optical conductivity can be obtained from the Kubo formula as follows
\bea\label{Kubo}
\sigma(\omega)=\frac{1}{i\omega}\langle J(\omega)J(-\omega)\rangle_{\rm retarded},
\eea
where $J(\omega)$ is the conserved current of the $U(1)$ global symmetry evaluated at zero spatial momentum. The right hand side can be calculated using the  AdS/CFT correspondence. Since we are working in the classical gravity regime, the leading contribution of the classical fluctuations of the gauge field is of ${\cal O}(N^2)$ which is due to the black hole background. As it is evident from the results of \cite{GR:2012}, the Fermi surfaces are positioned at ${\cal O}( N^0)$, therefore, to find the subleading order contributions of the Fermi surface to the conductivity, one needs to take into account the fermionic loop contributions to the gauge field propagator in the bulk (For more details see for example \cite{{ILM:2011},{Iqbal}}). In what follows, we will only consider the leading ${\cal O}(N^2)$ contribution.

To proceed, we consider small fluctuations of the gauge field in the $x_1$-direction, $\delta A_{x_1}\equiv a_1$. In general, turning on small fluctuations of the gauge field back reacts on the other fields of the model and we have to solve the equations of motion for the fluctuations of all them. More precisely one may have
\bea
A_M\rightarrow A_0 \delta_{M0}+a_M, \hspace{1cm}g_{MN}\rightarrow g_{MN}+h_{MN},\hspace{1cm}\alpha\rightarrow\alpha+\delta\alpha.
\eea

Since we are interested in calculating the two point function, it is enough to expand the action up to the quadratic level in fluctuations. To do so, it is convenient to first use the gauge freedom and set $h_{rM}=a_r=0$. Moreover in the level we are interested in, we may set  $\delta\alpha=0$. In this case the relevant quadratic terms of the action are obtained as follows\footnote{The subscripts ${(1)}$ and ${(2)}$ represent first and second order in metric fluctuations.}
\bea\label{Sem}
S_{\mathrm{EM}}^{(2)}\sim \int{d^{5}}x\, e^{4\alpha}\,\bigg[\frac{1}{2}E_r\bigg(\sqrt{-g}g^{tt}g^{rr}-\sqrt{-g}g^{tr}g^{tr}\bigg)_{(2)}E_r+\frac{1}{4}\sqrt{-g}f_{MN}f^{MN}\\-E_r\bigg(g^{tt}g^{rr}\sqrt{-g}\bigg)_{(1)}f_{tr}-\mathcal{Q}e^{-4\alpha}\bigg(h_t^i f_{ir}+h_r^i f_{ti}\bigg)\bigg],
\eea
where $E_r=\partial_r A_t$ and 
\be
f_{MN}=\partial_M a_N-\partial_N a_M,\;\;\;\;\;\;\;\;{\cal Q}=\sqrt{-g}e^{4\alpha} g^{tt} g^{rr}E_r.
\ee

For our purpose, we will focus on the zero momentum case which means we may set $a_1=a(r)e^{-i\omega t}$. In this case the gauge field fluctuations mix only with $h_{tx}$ component of the metric fluctuation. Therefore we find two coupled differential equations. Nevertheless one may eliminate $h_{tx}$ from the equations leading to the following differential equation for the gauge fluctuation\footnote{For a detailed discussion in the case of Einstein-Maxwell model  see Appendix 6.B of \cite{Iqbal}.}
\bea\label{Ax}
\partial_r\left[\sqrt{-g}e^{4\alpha}g^{rr}g^{xx}a'(r)\right]-\sqrt{-g}e^{4\alpha}g^{xx}\left(m_{eff}^2+\omega^2g^{tt}\right)a(r)=0,
\eea
where
\bea\label{Meff}
m_{eff}^2={\cal Q}^2e^{-4\alpha}g^{xx}g^{yy}g^{zz}.
\eea
Note that mixing between the gauge field and the graviton leads to a  mass term which preserves the gauge invariance.

For the case of two-charge black hole \eqref{ds2}, equation \eqref{Ax} reads 
%\beas
%m_{eff}^2=\frac{8\mu Q^2}{r^6(1+\frac{Q^2}{r^2})^{\frac{8}{3}}}=\frac{8(r_0^2+Q^2)^2Q^2}{r^{2/3}(r^2+Q^2)^{8/3}}.
%\eeas
\bea\label{gfeq}
\partial_r\left[\frac{(r^2-r_0^2)(r^2+r_0^2+2Q^2)}{r}a'(r)\right]-\left[\frac{8(r_0^2+Q^2)^2Q^2}{r(r^2+Q^2)^2}-\frac{\omega^2(r^2+Q^2)^2}{r(r^2-r_0^2)(r^2+r_0^2+2Q^2)}\right]a(r)=0.
\eea

Now the aim is to solve the above equation and find a solution with an ingoing boundary condition at the horizon. The corresponding retarded Green's function can
 then  be read from the asymptotic behavior of the solution near the boundary. Indeed near the boundary one finds
\be 
a\sim A+\frac{B}{r^2}
\ee 
and therefore
\be
\langle J_1(\omega)J_1(-\omega)\rangle_{\rm retarded}\sim \frac{B}{A}.
\ee

In order to find the retarded Green's function we utilize numerical method. To proceed we note that the equation \eqref{gfeq} has three free parameters, $Q$, $\omega$ and $r_0$. On the other hand since we are interested in conductivity at low energies, using the Kubo formula \eqref{Kubo}, for a given value of $Q$ we find the behavior of the retarded Green's function as a function of temperature (or $r_0$) for a small value of $\omega$. One may also find the behavior of the real part of the conductivity as a function of frequency at a fixed temperature\footnote{While we were at the last stage 
of our calculations, the paper \cite{DeWolfe:2012uv} appeared where the conductivity as a function of frequency for the dilatonic two-charge black hole  was been studied, numerically.}. 
Indeed using "NDSolve" in Mathematica the real part of the conductivity may be found as depicted in the Figure 1. 
\begin{figure}
\begin{center}
\includegraphics[scale=.8]{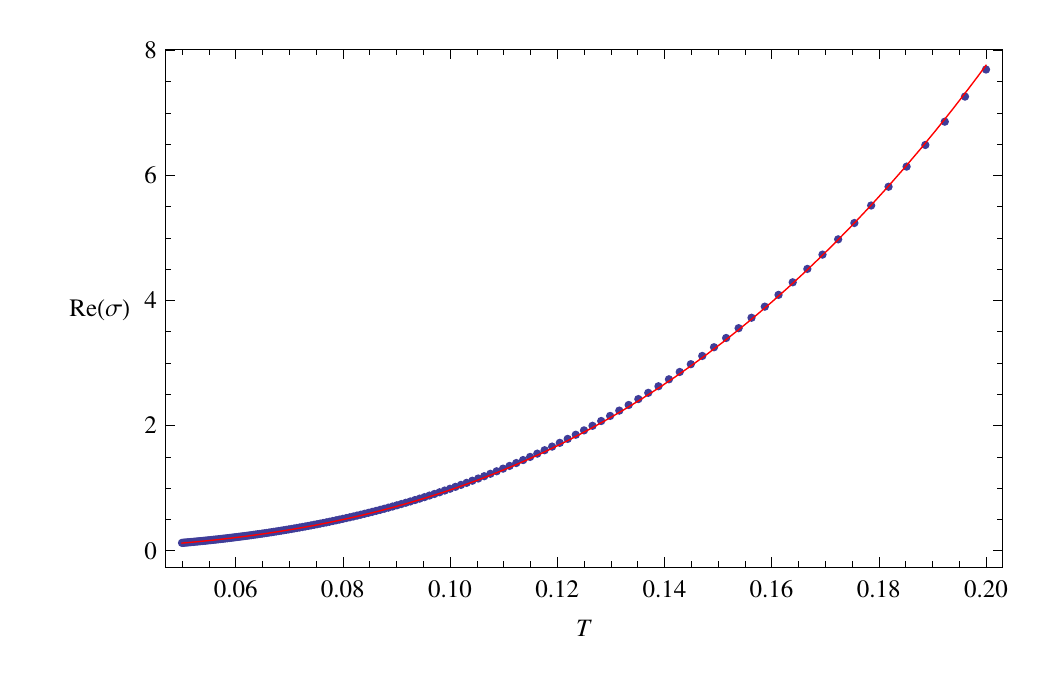}\label{sigmaT}
\includegraphics[scale=.9]{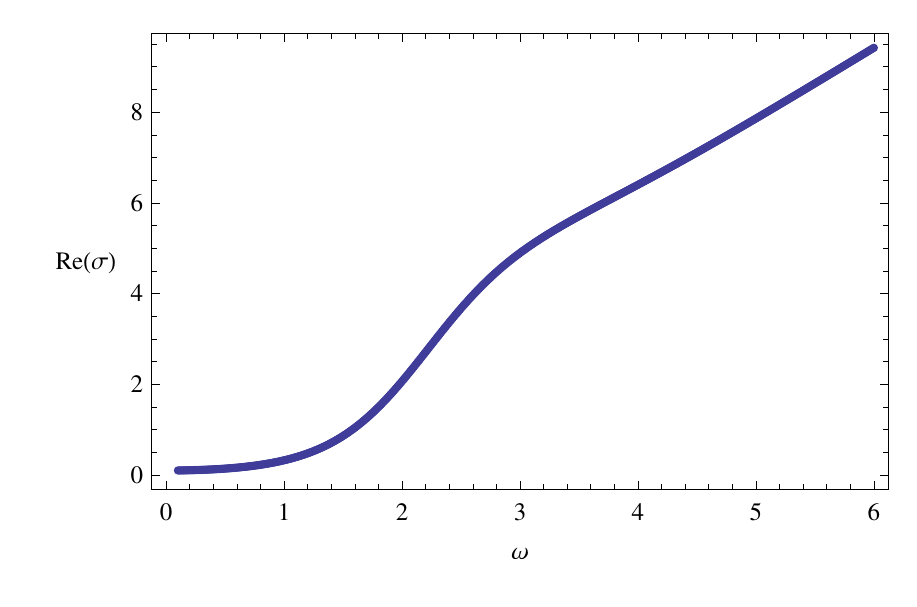}\label{sigmaW}
\caption{The (left) plot is the real part of conductivity versus temperature where we have set $Q=1$ and $\omega=0.001$. The numerical result is denoted by dots, while the red curve is the best fit which is ${\rm Re}(\sigma)\approx 0.97T^3$. Note that in order to make the resolution better we have rescaled Re($\sigma$) by factor of 1000. The (right) plot is the real part of conductivity versus frequency where we have set $Q=1$ and $T=\frac{1}{4\pi}$.}
\end{center}
\end{figure}
Using the numerical method the best fit one finds for the real part of the conductivity is
\be
{\rm Re}(\sigma)\approx 0.97\; T^3.
\ee

Alternatively one may find an analytic result for the conductivity as a function of temperature in the small frequency and small temperature limit (but with fixed $\omega/T$) by using the matching technique developed in \cite{FLMV:2009}. The procedure is as follows. Since the $g_{tt}$ component approaches zero as we approach the horizon (or taking the low energy limit, $\omega\rightarrow 0$), the $\omega$ dependent term in the equation of motion can not be considered as a small perturbation independent of how small $\omega$ could be. To overcome this problem we divide the geometry into two parts: inner and outer regions. These regions are defined as follows
\bea
\mathrm{Inner}&:&r=\frac{\omega}{2\xi}\;\;\;\;\;\;\mathrm{for}\;\;\;\;\;\;\epsilon<\xi<\xi_0=\frac{\omega}{2r_0}\\
\mathrm{Outer}&:&\frac{\omega}{2\epsilon}<r,
\eea  
while considering  the limit
\bea
\omega\to0,\hspace{4mm}\xi,\xi_0=\mathrm{finite},\hspace{4mm}\epsilon\to0,\hspace{4mm}\frac{\omega}{2\epsilon}\to0.
\eea

The inner and outer regions are described by $\xi$ and $r$ coordinates respectively. In these regions one may expand the solutions as follows
\bea
{a}_I(\xi)&=&{a}_I^{(0)}(\xi)+\omega\,{a}_I^{(1)}(\xi)+\cdots\label{AxIExpns}\\
{a}_O(r)&=&{a}_O^{(0)}(r)+\omega\,{a}_O^{(1)}(r)+\cdots\label{AxOExpns}.
\eea

It is then possible to solve the equation \eqref{gfeq} perturbatively and match the two solution in the overlapping region defined by $\xi\rightarrow 0$.

In the inner region substituting $r=\frac{\omega}{2\xi}$ and $r_0=\frac{\omega}{2\xi_0}$\footnote{We note that this is the same limit taken in the footnote \ref{foot} with $\lambda$ replaced by $\omega$.}, in the leading order of $\omega$ the equation \eqref{gfeq} reads 
\bea\label{LeadingIRax}
\xi\partial_\xi\left[\xi\left(1-\frac{\xi^2}{\xi^2_0}\right){a_{I}^{(0)}}'(\xi)\right]-\left[4-\frac{\xi^2}{\left(1-\frac{\xi^2}{\xi^2_0}\right)}\right]{a_{I}^{(0)}}(\xi)=0.
\eea

The most general solution of this equation satisfying the ingoing boundary condition at the horizon, up to a normalization factor, is 
\bea
{a_{I}^{(0)}}(\xi)=\xi^2(\xi^2-\xi_0^2)^u\,_2F_1\left(2+u,1+u,1+2u,1-\frac{\xi^2}{\xi_0^2}\right),
%&&\nonumber\\+b_I^{(0)}\,(\xi^2-\xi_0^2)^{-u}\,_2F_1\left(2-u,1-u,1-2u,1-\frac{\xi^2}{\xi_0^2}\right),\label{InAxSol}
\eea
where $u=-i\xi_0/2$ and $_2F_1$ is the hypergeometric function of the second kind. The solution near the matching region, $\xi\rightarrow 0$, up to an overall factor takes the form of
\bea\label{IRSol}
{a_{I}^{(0)}}(r)\simeq r^{2}+\frac{\mathcal{G}_a(\omega)}{16}r^{-2},
\eea
where $\mathcal{G}_a(\omega)$  is the retarded Green's function of  a conserved $U(1)$ current which is dual to the gauge field perturbation on the near horizon geometry
 \eqref{AdS2} that is calculated  in the Appendix. Note that to write the above equation we have utilized the relation between $\xi$ and $r$. 
 This is the expression to be matched to the outer region solution.

To solve the gauge field's fluctuation at the zeroth order in the outer region, we could simply set $\omega=0$, in equation \eqref{gfeq}, to get
\bea
\partial_r\left[\frac{1}{r}(r^2-r_0^2)(r^2+r_0^2+2Q^2){a_O^{(0)}}'(r)\right]-\left[\frac{8(r_0^2+Q^2)^2Q^2}{r(r^2+Q^2)^2}\right]a_0^{(0)}(r)=0.
\eea

The exact solution of the above equation is
\beas\label{axUVEq}
&&{a}_O^{(0)}(r)
=\frac{r^2}{r^2+Q^2}\bigg\{
a_O^{(0)}+\frac{b_O^{(0)}}{4}
\bigg[\frac{2Q^4}{r^2r_0^2(r_0^2+2Q^2)}+\frac{Q^2+r_0^2}{r_0^4}\log\left(1-\frac{r_0^2}{r^2}\right)\cr &&\cr && \;\;\;\;\;\;\;\;\;\;\;\;\;\;\;\;\;\;\;\;\;\;\;\;\;\;\;\;\;\;
-\frac{r_0^2+Q^2}{(r_0^2+2Q^2)^2}\log\left(1+\frac{r_0^2+2 Q^2}{r^2}\right)\bigg]\bigg\}.
\eeas

It is  easy to expand the solution around the matching region and match it with the equation \eqref{IRSol} to read $a_O^{(0)}$ and $b_O^{(0)}$ in terms of ${\cal G}_a(\omega)$. 
On the other hand expanding the solution near the boundary one finds
\bea
{a}_O^{(0)}(r)\bigg|_{r\to \infty}
&\simeq&
a_O^{(0)}\left\{1+\cdots\right\}+\left(-Q^2a_O^{(0)}-\frac{b_O^{(0)}}{2}\right)\left\{1+\cdots\right\}r^{-2}
\label{UVSolNB}\\
{a}_O^{(0)}(r)\bigg|_{\substack{r_0\to0\\r\to0}}
&\simeq&
\left(\frac{a_O^{(0)}}{Q^2}+\frac{b_O^{(0)}}{4Q^4}\right)\left\{1+\cdots\right\}r^2-\frac{b_O^{(0)}}{8}\left\{1+\cdots\right\}r^{-2}
\label{UVSolNH}
\eea
which leads to the leading order retarded Green's function (in frequency) as follows
\bea
G_{11}^R(\omega,T)=-Q^2\frac{8Q^4-\mathcal{G}_a(\omega)}{8Q^4+\mathcal{G}_a(\omega)}.
\eea

Using  the explicit expression of  $\mathcal{G}_a(\omega)$ given in the Appendix and plugging the resultant retarded Green's function in the Kubo formula \eqref{Kubo}, the conductivity at the leading order in frequency is obtained 
\bea
\sigma(\omega,T)=\frac{iQ^2}{\omega}+\frac{\pi^3}{Q^2}\,T^3+\cdots.
\eea 

The $T^3$ dependence is, indeed, what we expected from the numerical calculations. The first term is also what we would get in the classical gravity indicating that for a fixed temperature, the real part of the conductivity has a delta function behavior at $\omega=0$. It is worth to note that the delta function behavior is the artifact of simplifications in the gravity calculation which should be compared with a sample without impurities. Adding impurities would broaden the delta function into a Drude peak. From gravity point of view, this can be done by imposing a non-trivial boundary condition on the source of an operator at the boundary so that the translational invariance is broken. In this case, the Drude peak will appear from gravity calculations as well \cite{Tong:2012}.

%%%%%%%%%%%%%%%%%%%%%%%%%%%%%%%%%%%%%%%%%%%%%%%%%%%%%%%%%%%%%%%%%%%%%%%%%%%%%%
%%%%%%%%%%%%%%%%%%%%%%%%%%%%%%%%%%%%%%%%%%%%%%%%%%%%%%%%%%%%%%%%%%%%%%%%%%%%%%

\section{Entanglement Entropy}

In this section we will study the entanglement entropy of the dual theory by making use of the AdS/CFT correspondence. In this context  the holographic entanglement entropy can be, essentially,
 computed by minimizing a surface in the bulk gravity. More precisely, given a gravitational  theory with the  bulk Newton's constant $G_N$, the holographic entanglement entropy is given by \cite{RT:2006PRL,RT:2006}
\be
S_A=\frac{\mathrm{Area}(\gamma_A)}{4G_N},
\ee
where $\gamma_A$ is the minimal surface in the bulk whose boundary coincides with the boundary of the entangling region.

In what follows we will consider the extremal case where $r_0=0$. We note, however, that in this limit the metric \eqref{ds2}  develops a naked singularity. It is then interesting to see the effect of the singularity in the holographic entanglement entropy. To proceed  it is useful to define a new coordinate $z=1/r$ in which the metric of the extremal dilatonic black hole reads
\bea\label{ds(z)}
ds^2=f^{\frac{2}{3}}(z)\left[-h(z)dt^2+d\vec{x}^2+g(z)dz^2\right].
\eea   
where
\be
f(z)=\frac{1+z^2Q^2}{z^3},\;\;\;\;\;\;\;\;\;\;h(z)=\frac{1+2z^2Q^2}{(1+z^2Q^2)^2},\;\;\;\;\;\;\;\;\;\;g(z)=\frac{1}{1+2z^2Q^2}.
\ee

To begin, we will calculate entanglement entropy for a strip subsystem in the dual theory. From gravity point of view one needs to minimize a surface whose intersection with the boundary coincides to 
the strip.  The strip is defined by
\beas
-\frac{\ell}{2}\leq x_1\leq\frac{\ell}{2},\hspace{1cm}0\leq x_2,x_3\leq L.
\eeas
The profile of the surface in the bulk is also given by $x_1=x(z)$. Since the entanglement entropy will be calculated on equal time slices, the induced metric on the surface  reads
\bea
ds^2=f^{\frac{2}{3}}(z)\left[dx_2^2+dx_3^2+\left(g(z)+{x'}(z)^2\right)dz^2\right].
\eea   
Therefore the area is  
\bea
\mathcal{A}
&=&L^2\int_{\epsilon}^{z_*}dz\,f(z)\;\sqrt{g(z)+{x'}(z)^2}\label{A},
\eea
where $\epsilon$ is a UV cutoff and  $z_*$ is the 'turning point' where $\frac{dz}{dx}|_{z=z_*}=0$. The entanglement entropy is obtained by calculating $\mathcal{A}$ on its dominant trajectory. To extremize $\mathcal{A}$, one must use the variational principle for $x$. Actually, treating the area \eqref{A} as a one dimensional dynamical system and taking into account that it is independent of $x$, its momentum conjugate is a constant of motion 
\beas
f(z)\;\frac{x'}{\sqrt{{x'}^2+g(z)}}=f(z_*).
\eeas
It is then easy to find the turning point as a function of the strip length
\be\label{ll}
\ell=2\int_{0}^{z_*}dz\;\frac{f(z_*)}{f(z)}\;\sqrt{\frac{g(z)}{1-(f(z_*)/f(z))^2}}.
\ee
Finally the  entanglement entropy reads
\be
\mathcal{S}_{\mathrm{strip}}=
\frac{L^2}{4G_5} \int_{\epsilon}^{z_*}dz\;f(z)\;\sqrt{\frac{g(z)}{1-(f(z_*)/f(z))^2}}.\label{EEStrip}
\ee

We should now eliminate $z_*$ from the equations \eqref{ll} and \eqref{EEStrip} to find the entanglement entropy as a function of strip length $\ell$. Of course it is not an easy job to do that, though one may approximately estimate the leading order terms. To do so, we note that the main contribution to the entanglement entropy comes from the IR region which corresponds to the case where the turning point is deep in the bulk. In this case one may estimate the integrals \eqref{ll} and \eqref{EEStrip} for large $z_*$. In fact in leading order one finds
\be\label{IR-en}
%\ell\approx \frac{\pi}{\sqrt{2}Q}-\frac{2+\pi}{Q^2z_*},\;\;\;\;\;\;\;\;\mathcal{S}_{\mathrm{strip}}\approx \frac{H^2}{\epsilon}-\frac{3H^2Q}{8\sqrt{2}z_*}+\frac{5H^2}{32\sqrt{2}Qz_*^3},
\ell\approx \ell_0-\frac{\ell_1}{z_*},\;\;\;\;\;\;\;\;\mathcal{S}_{\mathrm{strip}}\approx \frac{L^2}{4G_5}\left(\frac{1}{2\epsilon^2}-\frac{S_1}{z_*}+\frac{S_3}{z_*^3}\right),
\ee
where $\ell_0,\ell_1,S_1$, and $S_3$ are numerical constant whose actual values depend on the order of expansion. It is important to note that $\ell_1$ and $S_1$ are always positive. Having found these expansion,  at the leading order one finds
\be
\mathcal{S}^{\mathrm{finite}}_{\mathrm{strip}}\approx\frac{L^2}{4G_5}\left[-\frac{S_1}{\ell_1}\left(\ell_0-\ell\right)+
\frac{S_3}{\ell_1^3}\left(\ell_0-\ell\right)^3\right].
\ee

\begin{figure}
\begin{center}
\includegraphics[scale=.8]{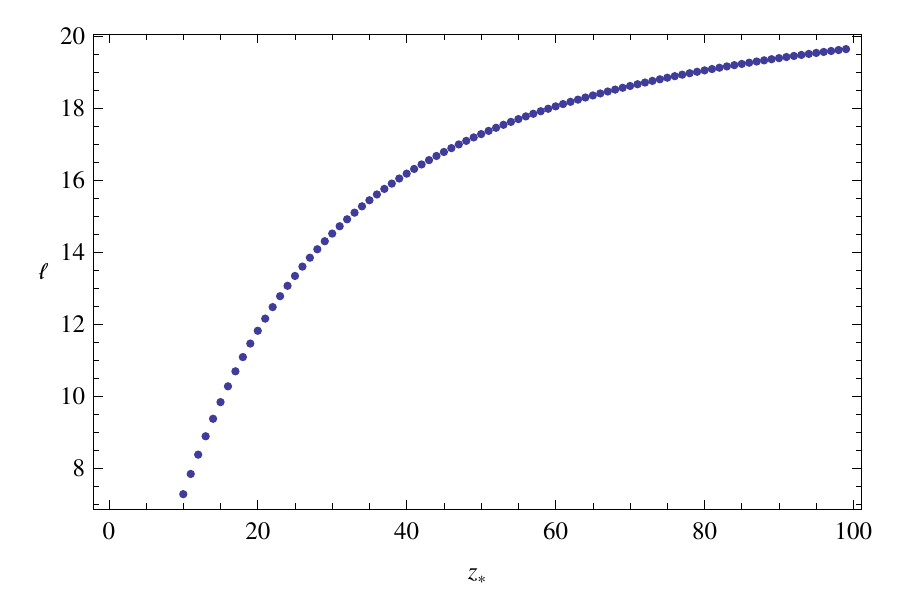}\label{EEstrp}
\includegraphics[scale=.8]{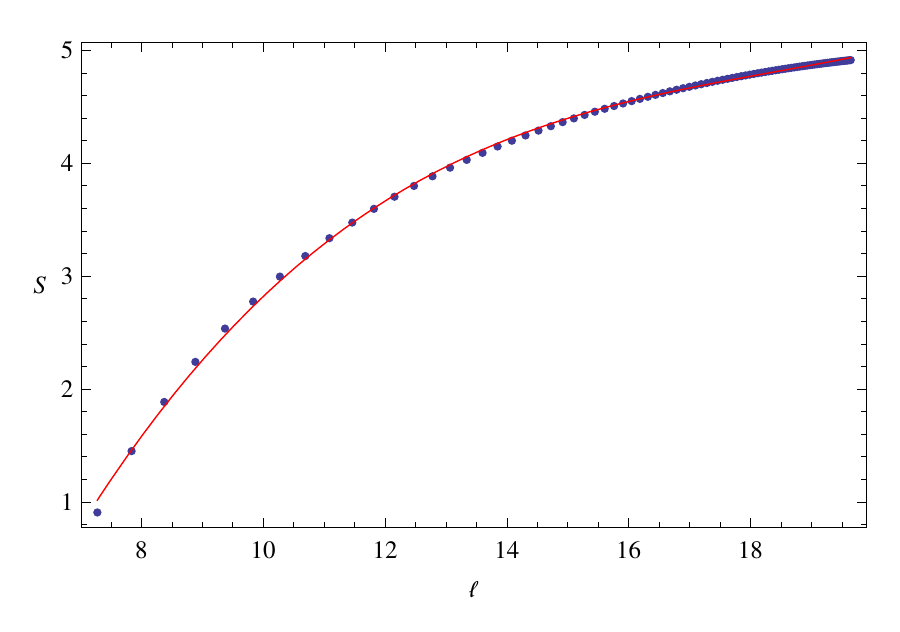}
\caption{Numerical results for the strip length as a function of the turning point (left) and the  finite part of entanglement entropy of a strip as a function of lenght $\ell$ (right).
The numerical data is ploted by dotes while the best fit is shown by the red curve. The plot of finite part of entanglement entropy is scaled by factor of 1000 for the
value of $Q=0.1$.  }
\end{center}
\end{figure}

It is also worth to solve the integrals in the equations \eqref{ll} and \eqref{EEStrip} numerically. For our numerical computations we set $Q=0.1$ and then using ``NIntegrate'' in  Mathematica  one can 
find $\ell$ as a function of turning point $z_*$ and also the finite part of the entanglement entropy
as a function of $\ell$, numerically. The results are shown in Figure \ref{EEstrp}. 

Using the numerical data the best fit for the  finite part of the entanglement entropy is
\be
\mathcal{S}^{\mathrm{finite}}_{\mathrm{strip}}\simeq \frac{L^2}{4G_5}\left( -9.2 000+ 2.0854 \ell - 0.10786 \ell^2+1.9514\times 10^{-3}\ell^3\right)\times 10^{-3}
\ee
which is in agreement with our approximated result.

In order to study the dependence of the entanglement entropy on the shape of the entangling region, it is elaborating  to consider a subsystem with other shape. In particular we will study the case of a circular subsystem.
% or  sphere 
To proceed we will  parametrize the metric of the three dimensional subspace $d\vec{x}^2$ as follows   
\bea
%&{\rm Cylinder}&:\;\;\;\;\;
d\vec{x}^2=d\rho^2+\rho^2d\theta^2+dx_3^2,\cr.
%&{\rm Sphere}&:\;\;\;\;\;d\vec{x}^2=d\rho^2+\rho^2(d\theta^2+\cos^2\theta d\phi).
\eea
Accordingly the subsystem is defined
\bea
%&{\rm Cylinder}&:\;\;\;\;\; 
0\le \rho\le\ell,\hspace{1cm}0\le\theta\le2\pi,\hspace{1cm}0\le x_3\le L,\cr
%&{\rm Sphere}&:\;\;\;\;\;0\le \rho\le \ell,\hspace{1cm}0\le\theta\le\pi,\hspace{1.23cm} 0\le \phi \le 2 \pi.
\eea
Therefore one finds
\bea
\mathcal{A}
&=&2\pi L \int_{\epsilon}^{z_*}dz\,f(z)\;\rho(z)\sqrt{\rho'(z)^2+g(z)}\label{Adisc},
\eea 
%$n=0,1$ for cylinder and sphere, respectively. 
where $z_*$ is again the turning point where $\rho'(r)$ diverges. The dominant trajectory is obtained by solving
\bea\label{discEoM}
\partial_z\left[f(z)\frac{\rho(z)\rho'(z)}{\sqrt{\rho'(z)^2+g(z)}}\right]=
f(z)\,\sqrt{\rho'(z)^2+g(z)},
\eea
with the boundary conditions $\rho(z\to0)=\ell$ and $\rho(z_*)=0$.

\begin{figure}
\begin{center}
\includegraphics[scale=.8]{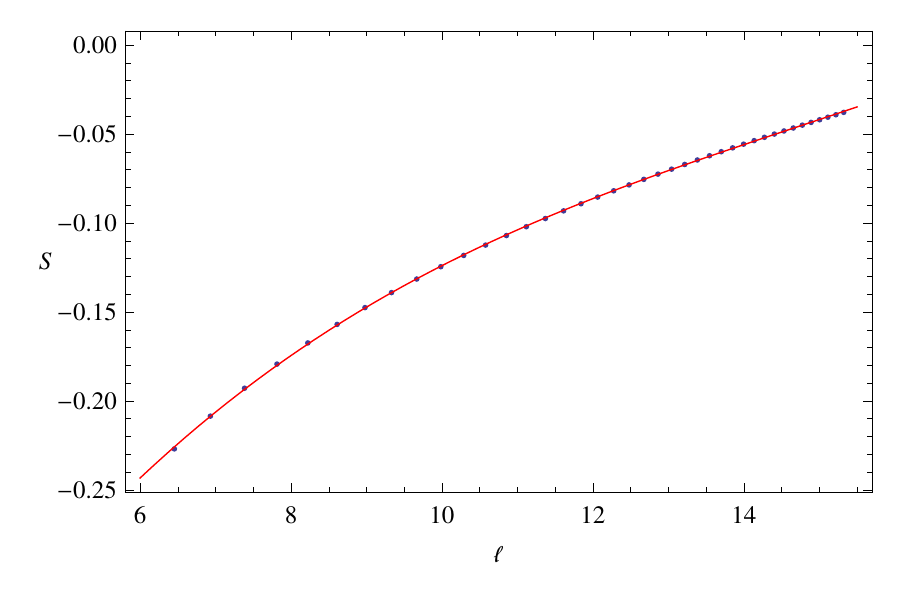}\label{EEcrc}
\caption{Numerical results for  the  finite part of entanglement entropy of a circular subsystem as a function of  length $\ell$. The numerical data is plotted by dotes while the best fit is shown by the red curve. Here we set  $Q=0.1$.}
\end{center}
\end{figure}

Unlike the strip case, in  the present case there is no conservation law, so one cannot find an explicit expression for $\rho'(z)$. Indeed  we will have to solve  both  \eqref{Adisc} and \eqref{discEoM} equations numerically. For this case the numerical result for the entanglement entropy as a function  of $\ell$ is shown in the Figure \ref{EEcrc}. In this case  the behavior of the entanglement entropy is well approximated by
\bea
%&{\rm Cylinder}&:\;\;\;\;\;
S^{\mathrm{finite}}_{\mathrm{cylinder}}\simeq \frac{\pi L}{2G_5}\left(-0.6147+0.8798\ell-0.005033\ell^2+0.000141\ell^3\right),\cr
%&{\rm Sphere}&:\;\;\;\;\;S^{\mathrm{finite}}_{\mathrm{sphere}}\;\simeq -5.2173-0.14128\ell+0.00972\ell^2+0.00005\ell^3.
\eea
It was shown in \cite{OTU:2011} that the logarithmic violation of the area law in the entanglement entropy corresponds to the existence of Fermi surfaces. We note, however, that 
since in the present case the Fermi surfaces are located at ${\cal O}( N^0)$, the entanglement entropy can not probe them. This is unlike the cases of hyperscaleing violating backgrounds where the existence of the Fermi surfaces may be observed by the violation of the area law of entanglement entropy at ${\cal O}(N^2)$ (for an analytic calculation of the entanglement entropy for general entangling region shapes on these backgrounds see \cite{HSS:2011}). 

It is also interesting to note that in the present case, unlike the RN AdS black hole at zero temperature, there is a bound for the strip length $\ell$. This can be seen both from the 
numerical result (see figure 2) as well as the IR analytic leading order contribution\eqref{IR-en}.
That means as the length $\ell$ exceeds the bound, there is no a minimal closed surface whose ends on the boundary coincides with the entangling region. Actually, it is very similar to the case where the entanglement entropy is computed on a background describing a confining phase of a system (see for example the review \cite{Nishioka:2009un}). 
We, note, whoever that in the confining phase the entanglement entropy approaches a constant number for large $\ell$, though in our case it has a power low behavior.

%%%%%%%%%%%%%%%%%%%%%%%%%%%%%%%%%%%%%%%%%%%%%%%%%%%%%%%%%%%%%%
%%%%%%%%%%%%%%%%%%%%%%%%%%%%%%%%%%%%%%%%%%%%%%%%%%%%%%%%%%%%%%

\section{More on the low energy behavior}
In the RN AdS$_5$ black hole, the IR geometry is AdS$_2\times\mathbb{R}^3$ and the scaling dimensions of the operators dual to the IR geometry depend on the charge of the dual fields. In general, depending on the charges and the momentum, the corresponding dimensions could be imaginary leading to an instability. This intermediate unstable phase, known as semi-local quantum liquid \cite{ILM:2012,ILM:2011},  may be modified to a Lifshitz geometry when the back-reaction coming from pair production are taken into account \cite{{Hartnoll:2010gu},{Hartnoll:2010ik},{Hartnoll:2011dm}}.   

For the case of the dilatonic two-charge black hole, the IR geometry is conformally AdS$_2\times\mathbb{R}^3$ and the presence of the conformal factor, indeed, alters the behavior of the fields in the IR region. In fact it is shown \cite{GR:2012} that the dimension of a charged spinor is always real and moreover the electric field vanishes in the near horizon limit. 
 In this section, in order to further explore  the IR behavior of the system we will study a charged scalar field on the dilatonic two-charge black hole. 

Consider a charged scalar field  minimally coupled to a gauge field 
\bea\label{ScalarAction}
S=- \int{d^{d+1}x \sqrt{-g}\left[g^{MN}\left(\partial_M +iqA_M\right)\phi^*\left(\partial_N -iqA_N\right)\phi+m^2\phi^* \phi\right]}.
\eea
The corresponding equation of motion is
\bea\label{ScalarEOM}
-\frac{1}{\sqrt{-g}}\partial_M \left(\sqrt{-g}\partial^M \phi\right)+q^2A^2\phi +iqA^M\partial_M \phi +\frac{iq}{\sqrt{-g}}\partial_M \left(\sqrt{-g}A^M \phi\right)+m^2\phi=0.
\eea
Assuming $\phi(t,x^i,r)=e^{-i\omega t+i\vec{k}.\vec{x}}\varphi(r)$ for a massless scalar field, the above equation on the dilatonic black hole geometry \eqref{ds2} in the extremal limit reads\footnote{Actually setting $k=0$ and $q=0$ this equation may be use to study the dynamical equation of the metric fluctuation ${h^x}_y$. Solving this equation and reading the two point function together with
 the corresponding Kubo formula lead to find the shear viscosity. In fact doing so we find that shear viscosity depends linearly on temperature.}
%\be\label{EOMphiG}
%-\frac{1}{\sqrt{-g}}\partial_r[\sqrt{-g}g^{rr}\phi'(\xi)]+[(\omega+qA_t)^2g^{tt}+k^2g^{ii}+m^2]\phi(r)=0.
%\ee
%In terms of $r$ one has
\bea\label{EOMphir}
r^{-1}\partial_r\left[r^3(r^2+2Q^2){\varphi}'(r)\right]+\left[\left(\omega-\frac{\sqrt{2}qQr^2}{(r^2+Q^2)}\right)^2\frac{(r^2+Q^2)^2}{r^2(r^2+2Q^2)}-k^2\right]{\varphi}(r)=0.
\eea

Now the aim is to solve this equation at low energy which may be used to  find the corresponding low energy retarded Green's function
of a scalar operator in the dual theory. To do so, we utilized the procedure introduced in \cite{{FLMV:2009},{FILMV:2011}} which we  have also used in section 2 for the gauge field fluctuation.
 Namely we will perturbativelly solve the above equation in inner and outer regions and match them in the matching region. The inner and and outer regions are defined by 
\bea
\mathrm{Inner}&:&r=\frac{\omega}{2\xi}\;\;\;\;\;\;\mathrm{for}\;\;\;\;\;\;\epsilon<\xi<\infty\cr
\mathrm{Outer}&:&\frac{\omega}{2\epsilon}<r,
\eea  
while considering  the limit
\bea
\omega\to0,\hspace{4mm}\xi=\mathrm{finite},\hspace{4mm}\epsilon\to0,\hspace{4mm}\frac{\omega}{2\epsilon}\to0.
\eea
The inner and outer regions are described by $\xi$ and $r$ coordinates respectively.  In these regions one may
expand the solutions as follows
\bea
{\varphi}_I(\xi)&=&{\varphi}_I^{(0)}(\xi)+\omega\,{\varphi}_I^{(1)}(\xi)+\cdots,\\
{\varphi}_O(r)&=&{\varphi}_O^{(0)}(r)+\omega\,{\varphi}_O^{(1)}(r)+\cdots\;.\label{PhiOExpns}
\eea
We will match these two solutions over the matching region defined by $\xi\rightarrow 0$.

At leading order in $\omega$-expansion the equation of motion for the scalar field in the inner region reduces to 
\bea
\xi^3\partial_\xi\left[\xi^{-1}{\phi_I^{(0)}}'(\xi)\right]+\left(\xi^2-\frac{k^2}{2Q^2}\right){\phi_I^{(0)}}(\xi)=0.
\eea
Note that  the gauge field does not contribute in the near horizon limit\footnote{Even if one had considered the mass term, it would not have  contributed to the near horizon dynamics, either.}. 
Imposing the ingoing boundary condition (the regularity condition) the most general solution of the above equation up to a normalization factor is
\bea\label{InSolPhi}
\phi_I^{(0)}(\xi)=\xi(J_\nu(\xi)+iY_\nu(\xi) ),\hspace{1cm}{\rm with}\;\;\nu\equiv\sqrt{1+\frac{k^2}{2Q^2}}.
\eea
Using the asymptotic behavior of the Bessel function,  near the matching region, $\xi\rightarrow 0$, up to an overall factor,  one finds
\bea\label{NBInPhi}
\phi_I^{(0)}(\xi)\bigg|_{\xi\to0}\simeq r^{-1+\nu}+\frac{\mathcal{G}_\phi(\omega)}{2^{2\nu}}r^{-1-\nu},
\eea
where $\mathcal{G}_\phi(\omega)$ is the retarded Green's function of an operator in the field theory dual to a charged scalar in the conformally AdS$_2$ geometry \eqref{AdS2} which is given in the Appendix. Note that since we always have $\nu>0$, the solutions never become degenerate. Also note that since we want to match this solution with that in outer region, we have written the above equation in terms of $r$ coordinate.

On the other hand in the outer region the leading contribution in $\omega$-expansion comes from the equation  \eqref{EOMphir} with $\omega=0$,
\bea
r^{-1}\partial_r\left[r^3(r^2+2Q^2){\varphi_O^{(0)}}'(r)\right]-\left[k^2-\frac{2q^2Q^2r^2}{(r^2+2Q^2)}\right]{\varphi_O^{(0)}}(r)=0,
\eea
whose general solution is
\bea
{\varphi_O^{(0)}}(r)=(r^2+2Q^2)^{\frac{q}{2}}\bigg[c^{(0)}_O\,r^{(-1+\nu)}\,_2F_1\left(\frac{q+3+\nu}{2},\frac{q-1+\nu}{2};1+\nu,\frac{-r^2}{2Q^2}\right)\nn\\+d^{(0)}_O\,r^{(-1-\nu)}\,_2F_1\left(\frac{q+3-\nu}{2},\frac{q-1-\nu}{2};1-\nu,\frac{-r^2}{2Q^2}\right)\bigg],
\eea
where $_2F_1$ is the hypergeometric function. Utilizing the asymptotic behaviors of the hypergeometric function $_2F_1$ one arrives at
\bea
\varphi^{(0)}_O\bigg|_{r\to0}&=&(2Q^2)^{\frac{q}{2}}\left[c^{(0)}_O\,\left\{1+\cdots\right\}\,r^{(-1+\nu)}+d^{(0)}_O\,\left\{1+\cdots\right\}\,r^{(-1-\nu)}\right],\label{NHOutPhi}
\cr
\varphi^{(0)}_O\bigg|_{r\to\infty}&=&\left(\alpha^{(0)}_\nu c^{(0)}_O+\alpha^{(0)}_{-\nu}d^{(0)}_O\right)\left\{1+\cdots\right\}+\left(\beta^{(0)}_\nu c^{(0)}_O
+\beta^{(0)}_{-\nu}d^{(0)}_O\right)\frac{1}{r^4}\left\{1+\cdots\right\},\label{NBOutPhi}
\eea
where $\alpha$ and $\beta$ are functions of $q,Q$ and $\nu$
\bea
&&\alpha^{(0)}_\nu=\frac{2}{Q^4}\eta_\nu,\cr &&\\ &&
\beta^{(0)}_\nu=\frac{\eta_\nu}{8}\bigg[8\left(q^3-q^2-q\nu^2-q\right)+\left(1-2\nu^2-2q^2+\left(\nu^2-q^2\right)^2\right)\cr &&\;\;\;\;\;\;\;\;\;
\;\;\;\;\;\;\;\;\;\;\;\;\;\;\;\;\;\;\;\;\;\;\;\;\;\;\;\;\;\;\;\;\;\;\;\;\;\;\;\;
\times\left(3-2\gamma-2\psi\left(\frac{-1-\nu-q}{2}\right)-2\psi\left(\frac{3-\nu+q}{2}\right)\right)\bigg]\nonumber
\eea
with
\bea
\eta_\nu&=&2^{\frac{1}{2} (-1+q+\nu )}Q^{(q+\nu-1)}\frac{\Gamma(1+\nu)}{\Gamma\left(\frac{3-q+\nu}{2}\right)\Gamma\left(\frac{3+q+\nu}{2}\right)}.
\eea
These expressions  can be used to read $c^{(0)}_O$ and $d^{(0)}_O$ by matching with the inner region behavior \eqref{InSolPhi}. Doing so, one finds
\be
c^{(0)}_O=(2Q^2)^{-\frac{q}{2}},\hspace{1cm}d^{(0)}_O=(2Q^2)^{-\frac{q}{2}}\frac{\mathcal{G}_\phi(\omega)}{2^{2\nu}}.
\ee
Having found these parameters, the retarded Green's function at leading order  reads
\bea
G_R(k,\omega)&=&\frac{\beta^{(0)}_\nu+2^{-2\nu}\beta^{(0)}_{-\nu}\mathcal{G}_\phi(\omega)}{\alpha^{(0)}_\nu +2^{-2\nu}\alpha^{(0)}_{-\nu}\mathcal{G}_\phi(\omega)}.
\eea
Going further one may also find higher order terms in the $\omega$ expansions in the above equation. The final result would have  the following form\cite{FLMV:2009}
\bea
G_R(k,\omega)&=&\frac{\beta_\nu+2^{-2\nu}\beta_{-\nu}\mathcal{G}_\phi(\omega)}{\alpha_\nu +2^{-2\nu}\alpha_{-\nu}\mathcal{G}_\phi(\omega)}.
\eea
where $\alpha$ and $\beta$ are 
\be
\alpha=\alpha^{(0)}+\alpha^{(1)}\omega+\alpha^{(2)}\omega^2+\cdots,\;\;\;\;\;\;\;\beta=\beta^{(0)}+\beta^{(1)}\omega+\beta^{(2)}\omega^2+\cdots\;,
\ee
and in principle all the coefficients $\alpha^{(i)}$ and $\beta^{(i)}$ can be calculated order by order.
 
It is interesting to compare this result to that for the RN AdS black hole studied in \cite{{FLMV:2009},{FILMV:2011}}. In the case of RN AdS black hole due to the contribution of the gauge field near the horizon, the scaling dimension $\nu$ could be imaginary. Therefore the coefficients $\alpha_\nu$ and $\beta_\nu$ could take imaginary values leading to an instability in the theory. 
In fact going from UV to IR the theory flows to an IR fixed point which is an unstable phase known as semi-local
quantum liquid. The instability in this phase is because the scalar becomes tachyonic and may condense which at zero spatial  momentum is similar to superconductor phase transition\cite{FLMV:2009}. It may also be understood from the fact that the IR limit of the model, being dual to AdS$_2$ gravity has non-zero entropy\footnote{For the case of fermions also $\nu_k\propto k$, thus pair production of fermions near the horizon does not distort the geometry \cite{GR:2012}.}.

On the other hand in our case, since in the near horizon limit the gauge field does not couple to  the scalar field, the scaling exponent $\nu$ is always a positive real number. As a result the coefficients $\alpha_\nu$ and $\beta_\nu$ are always real and the system is stable under scalar condensations\footnote{Changing the dilaton potential makes condensation possible\cite{Salvio:2012}.}. In other words it seems that as we go from UV to IR, the theory flows to a stable IR fixed point. This IR fixed point is gravitationally described by the gravity on a background which is conformally AdS$_2$ which  has vanishing entropy.

%%%%%%%%%%%%%%%%%%%%%%%%%%%%%%%%%%%%%%%%%%%%%%%%%%%%%%%%%%%%%%%%%
%%%%%%%%%%%%%%%%%%%%%%%%%%%%%%%%%%%%%%%%%%%%%%%%%%%%%%%%%%%%%%%%%

\section{Conclusions}
In this paper we have studied certain holographic aspects of two-charge dilatonic AdS$_5$ black hole. This model has been proposed to describe a holographic Fermi liquid model  which could be in same
 universality class of the Landau Fermi liquid \cite{GR:2012}. The fermionic properties of this model  were studied in \cite{GR:2012} by probing the background by a massless fermion, while in the present paper we considered different bosonic aspects. In particular we have studied the entanglement entropy and conductivity. We have also studied a charged scalar probing the 
geometry where we observed that the theory flows to a stable fixed point described by a geometry which is conformally 
AdS$_2\times\mathbb{R}^3$.  

Combining the results of \cite{GR:2012} and those in the present work we may summerize the properties of the model as follows

\begin{itemize}
\item There are several exact information about different quantities in the model such as Fermi surface positions.
\item The theory flows to an IR fixed point which seems to be stable against fermions pair production and also boson condensation.
\item The IR geometry which is conformally AdS$_2\times\mathbb{R}^3$ has zero entropy.
\item The specific heat, entropy and shear viscosity are linear in temperature while the conductivity 
goes as $T^3$, at low energies and low temperature.
\end{itemize}     

We note, however,  that the background has a naked singularity in its extremal limit. Moreover, even though the background may be uplifted to type IIB supergravity, the considerations of the present work and those in \cite{GR:2012} are still bottom-up approach. It would be interesting to find a top-down model enjoying such properties (see for example \cite{{Gauntlett:2011mf},{DeWolfe:2012uv}}). 

%%%%%%%%%%%%%%%%%%%%%%%%%%%%%%%%%%%%%%%%%%%%%%%%%%%%%%%%%%%%%%%%%%
%%%%%%%%%%%%%%%%%%%%%%%%%%%%%%%%%%%%%%%%%%%%%%%%%%%%%%%%%%%%%%%%%%

\section*{Acknowledgments}

We would like to thank D. Allahbakhshi, H. Ebrahim, N. Iqbal, A. E. Mosaffa, A. Vaezi for useful discussions. This work is supported by Iran National Science Foundation (INSF).

%%%%%%%%%%%%%%%%%%%%%%%%%%%%%%%%%%%%%%%%%%%%%%%%%%%%%%%%%%%%%%%%%%
%%%%%%%%%%%%%%%%%%%%%%%%%%%%%%%%%%%%%%%%%%%%%%%%%%%%%%%%%%%%%%%%%%

\section*{Appendix}
In this appendix we discuss retarded Green's functions for operators in a field theory whose gravity dual are a gauge field and a charged scalar in the conformally AdS$_2\times\mathbb{R}^3$ background \eqref{AdS2}. We note that there is also a non-zero gauge field which in the near extremal limit is
\be
A_\tau=-\frac{1}{2\sqrt{2}Q\xi^2}\left(1-\frac{\xi^2}{\xi_0^2}\right).
\ee
Of course as it has been pointed out in \cite{GR:2012} it  plays no role in what follows.

\subsection*{Gauge field}
Consider small fluctuations of the gauge field in $x$-direction on the near extremal background \eqref{AdS2}. The corresponding equation of motion can be read from  \eqref{gfeq} 
\bea\label{IRsax}
\xi\partial_\xi\left[\xi\left(1-\frac{\xi^2}{\xi^2_0}\right)a'_x(\xi)\right]-\left[4-\frac{\omega^2\xi^2}{\left(1-\frac{\xi^2}{\xi^2_0}\right)}\right]a_x(\xi)=0.
\eea
The most general solution satisfying the ingoing boundary condition at the horizon, up to a normalization constant, is
\bea
{a_{x}}(\xi)=\xi^2(\xi^2-\xi_0^2)^{-i\xi_0\omega/2}\,_2F_1\left(2-\frac{i{\omega}\xi_0}{2},1-\frac{i{\omega}\xi_0}{2},1-i{\omega}\xi_0,1-\frac{\xi^2}{\xi_0^2}\right)
%&&\nonumber\\+\mathcal{B}\,(\xi^2-\xi_0^2)^{-u}\,_2F_1\left(2-u,1-u,1-2u,1-\frac{\xi^2}{\xi_0^2}\right).\label{axIRSol}
\eea
Using the asymptotic behavior of the hypergeometric function near $\xi\rightarrow 0$ one finds
\bea
{a_{x}}(\xi)\bigg|_{\xi\to0}\simeq \frac{A(\omega)}{\xi^{2}}+B(\omega)\xi^{2}
\eea
where
\bea
A(\omega)&=&\frac{\xi_0^4(-\xi_0)^{-i\xi_0\omega}\Gamma(1-i\xi\omega)}{\Gamma\left(1-\frac{i\xi_0\omega}{2}\right)\Gamma\left(2-\frac{i\xi_0\omega}{2}\right)},\nonumber\\
B(\omega)&=&\frac{\omega A(\omega)}{16\xi_0^3}\bigg[4i+2\xi\omega+2i\xi_0^2\omega^2\cr \nonumber\\&&
-\left(\xi_0\omega+\frac{\xi_0^3\omega^3}{4}\right)\left(-3+4\gamma+2\psi\left(1-\frac{i\xi_0\omega}{2}\right)+
2\psi\left(2-\frac{i\xi_0\omega}{2}\right)\right)\bigg],\nonumber
\eea
where $\gamma$ is Euler number and $\psi(x)$ function is defined by $\psi(x)=\frac{d\log\Gamma(x)}{dx}$. From these expressions the retarded Green's function at leading order in $\omega$ is 
\bea\label{IRG}
\mathcal{G}_a(\omega)\approx\frac{i\omega}{4\xi_0^3}=2i\omega(\pi T)^3.
\eea
\subsection*{Scalar field}
Consider a charged scalar field in the near horizon geometry of the extremal dilatonic two-charge black hole. The corresponding metric is given 
by \eqref{AdS2} for $\xi_0\rightarrow \infty$. The equation of motion of the scalar field is
\bea\label{InPhi}
\xi^3\partial_\xi\left[\xi^{-1}\phi'(\xi)\right]+\left(\omega^2\xi^2-\frac{k^2}{2Q^2}\right)\phi(\xi)=0.
\eea
It is important to note that in the present case the electric field approaches zero in the near horizon limit \cite{GR:2012} and therefore it has no contribution in the above equation.

The general solution to the above equation is
\bea
\phi(\xi)=\xi\left[\mathcal{C}J_\nu(\omega\xi)+\mathcal{D}Y_\nu(\omega\xi)\right],\hspace{1cm}\nu\equiv\sqrt{1+\frac{k^2}{2Q^2}}.
\eea
Imposing the ingoing boundary condition (a proper regularity condition at  $\xi\to\infty$) one gets $\mathcal{D}=i\mathcal{C}$. Moreover from the near boundary ($\xi\to0$) analysis of the equation \eqref{InPhi} one  finds that  independent solutions on the boundary are
\bea
\phi(\xi)\bigg|_{\xi\to0}\sim\xi^{1\pm\nu}.
\eea
Reading off the coefficients from the exact solution, near the boundary one arrives at
\bea
\phi(\xi)\bigg|_{\xi\to0}\simeq c_I\xi\left[\frac{1+i\cot{\pi\nu}}{\Gamma(1+\nu)}\left(\frac{\omega\xi}{2}\right)^\nu-\frac{i}{\Gamma(1-\nu)\sin{\pi\nu}}\left(\frac{\omega\xi}{2}\right)^{-\nu}\right],
\eea
and thus the retarded Green's function reads
\bea
\mathcal{G}_\phi(\omega)=-e^{-i\pi\nu}\frac{\Gamma(1-\nu)}{\Gamma(1+\nu)}\left(\frac{\omega}{2}\right)^{2\nu}.
\eea
It is important to note that the scaling exponent $\nu$ is always a positive real number. Therefore unlike the RN AdS case, the system should be stable. It is worth to recall that in the RN AdS case due to the contribution of the electric field the scaling exponent could be  imaginary leading to an instability. This is because  in this csae the scalar field on the AdS$_2$ geometry becomes tachyonic. 

In the near extremal case the  equation of motion of the scalar field is
\bea
\xi^3\partial_\xi\left[\xi^{-1}\left(1-\frac{\xi^2}{\xi^2_0}\right)\phi(\xi)\right]+\left[\frac{\omega^2\xi^2}{\left(1-\frac{\xi^2}{\xi^2_0}\right)}-\frac{k^2}{2Q^2}\right]\phi(\xi)=0,
\eea
whose general solution is
\bea
\phi&=&C_1 \xi^{1-\nu}(\xi^2-\xi_0^2)^{i\xi_0\omega/2}\,_2F_1\left(\frac{i{\omega}\xi_0}{2}+\frac{1}{2}-\frac{\nu}{2},
\frac{i{\omega}\xi_0}{2}+\frac{1}{2}-\frac{\nu}{2},1-\nu,\frac{\xi^2}{\xi_0^2}\right)\cr &&\cr &+&
C_2 \xi^{1+\nu}(\xi^2-\xi_0^2)^{i\xi_0\omega/2}\,_2F_1\left(\frac{i{\omega}\xi_0}{2}+\frac{1}{2}+\frac{\nu}{2},
\frac{i{\omega}\xi_0}{2}+\frac{1}{2}+\frac{\nu}{2},1+\nu,\frac{\xi^2}{\xi_0^2}\right),
\eea
where $\nu^2=1+\frac{k^2}{2Q^2}$. Imponsing ingoing boundary condition at the horizon one finds a relation between $C_1$ and $C_2$ which can be
used to obtain the retarded Green's function as follows
\be
{\cal G}_\phi(\omega)=-(2\pi T)^{-2\nu}\frac{\Gamma(1-\nu)}{\Gamma(1+\nu)}\;
\frac{\Gamma^2(\frac{1-i\omega/2\pi T+\nu}{2})}{\Gamma^2(\frac{1-i\omega/2\pi T-\nu}{2})}.
\ee

\end{document}